\documentclass[prb,twocolumn,floatfix,showpacs,preprintnumbers, 
               superscriptaddress,amsmath,amssymb]{revtex4}
\usepackage[dvips]{graphics}
\usepackage{psfrag}
\usepackage{graphicx}
\usepackage{epsf}
\usepackage{ucs}
\usepackage[utf8x]{inputenc}
\usepackage{amssymb}
\newcommand{\be}{\begin{eqnarray}}
\newcommand{\ee}{\end{eqnarray}}

\begin{document}
\title{Influence of vibrational modes on the  
electronic properties of DNA }

\author {Benjamin B. Schmidt} 
\affiliation{Institut f\"ur Theoretische Festk\"orperphysik
and DFG-Center for Functional Nanostructures (CFN),
Universit\"at Karlsruhe, 76128 Karlsruhe, Germany}
\affiliation{Forschungszentrum Karlsruhe, Institut f\"ur Nanotechnologie,
  Postfach 3640, 76021 Karlsruhe, Germany}
\author {Matthias H. Hettler} 
\affiliation{Forschungszentrum Karlsruhe, Institut f\"ur
  Nanotechnologie, Postfach 3640, 76021 Karlsruhe, Germany}
\author {Gerd Sch\"on} 
\affiliation{Institut f\"ur Theoretische Festk\"orperphysik
and DFG-Center for Functional Nanostructures (CFN),
Universit\"at Karlsruhe, 76128 Karlsruhe, Germany}
\affiliation{Forschungszentrum Karlsruhe, Institut f\"ur
  Nanotechnologie, Postfach 3640, 76021 Karlsruhe, Germany}

\date{\today}

%%%%%%%%%%%%%%%%%%%%%%%%%%%%%%%%%%%%%%%%%%%%%%%%%%%%%%%%%%%%%%%%%%%%%%%%%%%%%%
\begin{abstract}
We investigate the electron (hole) transport 
through short double-stranded DNA wires
in which the electrons are strongly coupled to the specific 
vibrational modes (vibrons) of the DNA. We analyze the problem
starting from a tight-binding model of DNA, 
with parameters derived from ab-initio calculations, and
describe the dissipative transport by equation-of-motion techniques. 
For homogeneous DNA sequences
like Poly- (Guanine-Cytosine) we find the transport to be quasi-ballistic 
with an effective density of states which is modified by the 
electron-vibron coupling. At low temperatures the linear conductance 
is strongly enhanced, but above the `semiconducting' gap it is affected much less. In contrast, for inhomogeneous (`natural') 
sequences almost all states are strongly localized, and transport 
is dominated by dissipative processes. In this case, a non-local 
electron-vibron coupling influences the conductance in a qualitative
and sequence-dependent way. 
\end{abstract}

\maketitle
%%%%%%%%%%%%%%%%%%%%%%%%%%%%%%%%%%%%%%%%%%%%%%%%%%%%%%%%%%%%%%%%%%%%%%%%%%%%%%

\section{Introduction} 
Transport measurements on DNA display a wide range of properties,  
depending on the measurement setup, the environment, and
the specific molecule, with behavior ranging from insulating 
\cite{Braun98} via semi-conducting \cite{Porath00} to quasi-metallic 
\cite{Xu04}. The variance of experimental results as
well as ab-initio calculations \cite{Starikov05}  suggest that the 
environment and its influence via the vibrational modes (vibrons) of 
DNA  are an 
important factor for the electronic transport properties of DNA wires.

Numerous recent articles addressed the electronic properties 
in a microscopic 
approach \cite{Asai04,Galperin05a,Gutierrez05b,Gutierrez06}. 
Typically, the DNA is described within a
tight-binding model for the electronic degrees of freedom with
parameters either taken from ab-initio quantum chemistry
simulations \cite{Voityuk00,Starikov05,Senthilkumar05} 
or motivated by a fit to experiments. The variance of 
qualitatively different tight-binding models is large, 
ranging from involved all-atomic representations to models where each 
base pair is represented by only a single orbital. 

Several suggestions in the past 
stressed the importance of the environment and vibrational modes on the
electron transfer \cite{Conwell00,Henderson99} and transport \cite{Yoo01,Troisi02}.
However, the vibrons have been treated so far 
only within very simple models, where specifically only a local
electron-vibron coupling has been taken into account \cite{Gutierrez05b}.
If the coupling is sufficiently strong, this leads to the formation of
polarons, i.e., a bound state of an electrons with a 
a lattice distortion. While these approaches are sufficient
to describe the transition from elastic (quasi-ballistic) to 
inelastic (dissipative) transport they ignore the fact that the non-local
electron-vibron coupling strength is comparable in 
magnitude to the local one \cite{Starikov05}. 
Furthermore, as the non-local electron-vibron coupling leads
effectively to a vibron-assisted hopping, the proper inclusion of this
coupling can be important for transport through the inhomogeneous sequences of
`natural' DNA.   

In this paper we study electron transport through double-stranded 
DNA wires strongly coupled, both locally and non-locally, to 
vibrational modes of the DNA. 
The DNA base pairs are represented by single tight-binding orbitals,
with energies differing for Guanine-Cytosine (GC)
and Adenine-Thymine (AT) pairs. The vibrational modes are also 
coupled to the surrounding environment (water or buffer solution) which we 
represent by a harmonic oscillator bath. This extension allows for
dissipation of energy and opens the possibility of
inelastic transport processes. We address the influence of specific DNA 
vibrational modes on transport in the frame of
equation-of-motion techniques, with parameters motivated by ab-initio 
calculations \cite{Starikov05,Senthilkumar05}. 

Our two main results are: 
1) For homogeneous DNA sequences
like Poly- (Guanine-Cytosine) wires the vibrons strongly enhance the
linear conductance at low temperatures. At large bias the vibrons affect 
the conductance only weakly, which remains dominated by quasi-ballistic 
transport through extended electronic states. 
2) For inhomogeneous (`natural') 
sequences almost all states are strongly localized, and transport 
is dominated by inelastic (dissipative) processes. 
In this case, the presence of a non-local electron-vibron coupling, 
leading to `vibron-assisted' electron hopping', influences the conductance 
in a qualitative and quantitative way.

The paper is organized as follows: in the following section 
we introduce the model and sketch briefly the techniques used 
to derive the transport properties.  In Section~\ref{subsec:homo} 
we present our results for homogeneous  DNA wires, while in 
Section~\ref{subsec:inhomo} we discuss a specific inhomogeneous 
DNA sequence, that has been studied in recent experiments.
A summary is provided in  Section~\ref{sec:summary}.
Details of the applied technique can be found in the  
Appendix  \ref{sec:appendix}.

\section{Model and technique}
\label{sec:model_and_technique}
Quantum chemistry calculations \cite{Artacho03,Lewis03}
show that the highest occupied molecular orbital (HOMO)
of a DNA base pair is located on the Guanine or Adenine,
whereas the lowest unoccupied molecular orbital (LUMO) is located
on the Thymine and Cytosine. Between HOMO and LUMO there is an 
energetic gap of approximately $2-3\, e \rm{V}$. Experimental
evidence hints to the prevalence of hole transport through DNA.
Given the energetic and spatial 
separation of HOMO and LUMO and considering sufficiently low bias
voltage we can represent in a minimal model one base pair by a single
tight-binding orbital.

We consider a DNA sequence with N base pairs, the first and last of which
are coupled to semi-infinite metal electrodes. 
We further allow for a coupling to (in general multiple) vibrational modes, that 
can be excited by local and non-local coupling to the charge carriers on the DNA. 
These modes in turn are coupled to the environment. When later performing the numerical 
calculations we will restrict ourselves to a single vibrational mode of
the DNA base pair, e.g., the `stretch' mode \cite{Starikov05}.

We thus arrive at the Hamiltonian
$H=H_{\rm el}+H_{\rm vib}+H_{\rm el-vib}+H_{\rm L}+H_{\rm R}+H_{\rm T,L}+H_{\rm T,R}+H_{\rm bath}$
with
\be
H_{\rm el} &=& \sum_i \epsilon_i a_i^{\dagger}a_i -\sum_{i,j; i \neq j} t_{ij} a_i^{\dagger}a_j \nonumber\\
H_{\rm T,L}+H_{\rm T,R} &=& \sum_{n,k,i} \left[ t_{in}^{r}c_{nr}^{\dagger}a_i+t_{in}^{r*}a_i^{\dagger}c_{n r} \right]\nonumber\\
H_{\rm vib}&=& \sum_{\alpha} \omega_{\alpha} B_{\alpha}^{\dagger}B_{\alpha} \nonumber\\
H_{\rm el-vib} &=& \sum_{\alpha} \sum_i \lambda_{0} \, a_i^{\dagger}a_i (B_{\alpha}+B_{\alpha}^{\dagger})\nonumber\\
&+& \sum_{\alpha} \sum_{i,j;i \neq j} \lambda_{ij} \, a_i^{\dagger}a_j (B_{\alpha}+B_{\alpha}^{\dagger}) \; .
\ee
The index $r=\rm L,R$ describing left and right electrode. 
The term $H_{\rm el}$ describes the electrons in the DNA chain with
operators $a_{i}^{\dag}, a_{i}$ in a single-orbital
tight-binding representation with on-site energies 
$\epsilon_i$ of the base pairs and hopping $t_{ij}$ between 
neighbouring base pairs.
Both on-site energies and hopping depend on the base pair 
sequence, e.g., the on-site energy of a Guanine-Cytosine base
pair differs from the on-site energy of a Adenine-Thymine base pair.
For the hopping matrix elements $t_{ij}$ we used the values calculated
by Siebbeles et al.\cite{Senthilkumar05} who studied intra- and 
inter-strand hopping between the bases in DNA-dimers by density functional theory (DFT).
They computed direction-dependent values for all possible hopping
matrix elements in such dimers. Adapting these results to our
simplified model of base pairs we obtain the hopping elements 
denoted in table~\ref{tab:hopping} \, \footnote{We assume that the
  holes can only reside on the purine bases, G or A. 
The hopping integrals between two purines depend
on the specific bases involved and to which strands these two bases
belong to. The values we used, were the values for the hopping integral
$J$ of the first, second and fifth row in the table 3 of Ref. 10. 
They are exactly reproduced in our table 1.}.
\begin{table}
5'-XY-3'(all in $e$V)\\
\begin{tabular}{|c|c|c|c|c|}\hline 
X$\diagdown$ Y & G  & C  & A  & T \\ \hline 
G & 0.119  & 0.046  & -0.186  & -0.048 \\ \hline 
C &  -0.075 & 0.119  & -0.037  & -0.013 \\ \hline 
A & -0.013 & -0.048  & -0.038  & 0.122 \\ \hline 
T & -0.037 & -0.186  & 0.148  & -0.038 \\ \hline 
\end{tabular}
\caption{Hopping integrals $t_{ij}$ taken form Ref.~\onlinecite{Senthilkumar05} 
and adapted to our model. The notation 5'-XY-3' indicates the direction
along the DNA strand (see e.g. Fig. 1b in Ref.~\onlinecite{Endres04}.) \label{tab:hopping}}
\end{table}
The number in the G row and the A column denotes the hopping matrix element
from a GC base pair to an AT base pair to its `right' (to the 3'
direction), for example. 

The terms $H_{\rm L/R}$ refer to the left and right electrodes. 
They are modeled by non-interacting electrons, 
described by  operators 
$c_{k\, \rm L/R}^{\dag}, c_{k\,\rm  L/R}$,
with a flat density of states $\rho_e$ (wide band limit).
The chemical details of the coupling between the DNA and the electrodes
are not the focus of this work. For our purposes it is fully characterized by 
$H_{\rm T,L}+H_{\rm T,R}$, which leads to a level broadening of the base pair 
orbitals coupled to the electrodes characterized by the linewidths 
$\Gamma^{\rm L}$ and $\Gamma^{\rm R}$.

The vibronic degrees of freedom are described by $H_{\rm vib}$, 
with bosonic operators $B_{\alpha}$ and $B_{\alpha}^{\dagger}$ for the vibron mode 
with frequency $\omega_{\alpha}$. $H_{\rm el-vib}$ couples the electrons
on the DNA to the vibrational modes, 
where $\lambda_{0}$ and $\lambda_{ij}$ are the strengths for the 
local and non-local electron-vibron coupling,
respectively. We further restrict the non-local coupling terms 
to nearest neighbors, $\lambda_{ij} = \lambda_{1}
\delta_{i,j=i \pm 1}$.
Note that the vibron modes and their coupling to electrons is assumed
independent of the base pairs involved, 
an approximation that is reasonable for some modes of interest, 
including the base pair stretch mode \cite{Starikov05}.
The strength of the electron-vibron coupling for various vibrational
modes has been computed in Ref.~\onlinecite{Starikov05} for 
homogeneous dimers and tetramers of AT and GC pairs. 
Here we consider also inhomogeneous sequences, 
for which the electron-vibron couplings are not known. As a model we take
$\lambda_{0}$ and $\lambda_{1}$ as parameters, independent of the base 
pairs involved, for which we choose values in rough agreement with 
estimates for the `stretch' mode of Ref.~\onlinecite{Starikov05}. 
This should be sufficient for a qualitative discussion of the 
effects that arise from the electron-vibron coupling in DNA.

The vibrons are coupled to the environment, the 
microscopic details of which do not matter. We model it by a 
harmonic oscillator bath $H_{\rm bath}$, whose 
relevant properties are summarized by its linear (`Ohmic') power spectrum
(or spectral function) up to a high-frequency cut-off $\omega_c$\cite{Weiss99}.
The coupling of the vibrons to the bath changes the vibrons
spectra from discrete (Einstein) modes to continuous spectra with
a peak around the vibron frequency. Physically, the
coupling to a bath allows for dissipation of electronic and vibronic
energy. This dissipation is crucial for the stability of the DNA molecule
in a situation where inelastic contributions to the current 
dissipate a substantial amount of power on the DNA itself. 

As mentioned before we only consider a single vibrational mode when performing 
the numerical calculations. This vibrational mode with resonance frequency $\omega_0$
coupled to the bath is then described by a spectral density 
\begin{align}
D(\omega)=\frac{1}{\pi}\left(
  \frac{\eta(\omega)}{(\omega-\omega_0)^2+\eta(\omega)^2}-
\frac{\eta(\omega)}{(\omega+\omega_0)^2+\eta(\omega)^2}\right)\, ,
\end{align}
with a frequency dependent broadening $\eta(\omega)$ which arises from the 
vibron-bath coupling. For the 'Ohmic' bath with weak 
vibron-bath coupling and cut-off $\omega_c$ we consider $\eta(\omega)=0.05\, \omega \, \theta(\omega_c-\omega)$.
Mathematically the crossover from the discrete vibrational modes to a continuous 
spectrum of a single mode is done by substituting 
$\sum_{\alpha} \delta(\omega-\omega_{\alpha}) \rightarrow \int d\omega D(\omega)$.

For the strong electron-vibron coupling predicted for DNA\cite{Starikov05},
one expects polaron formation, with a polaron size of a few base pairs. 
To describe these polarons (a combined electron-vibron `particle')
theoretically we apply the Lang-Firsov unitary 
transformation with the generator function $S$ to our Hamiltonian
(see e.g. Ref.~\onlinecite{Mahan})
\be
\bar{H}=e^{S}He^{-S} \; ;  \; S=-\sum_{i \alpha}
\frac{\lambda_{0}}{\omega_{\alpha}} \, a_i^{\dagger}a_i
\left[B_{\alpha}-B_{\alpha}^{\dagger} \right]\, .
\ee
After introducing transformed electron and vibron operators according to
\be
\bar{a_i}  &=&  a_i\chi\\
\bar{B_{\alpha}} &=& B_{\alpha}-\sum_i \frac{\lambda_{0}}{\omega_{\alpha}} \, a_i^{\dagger}a_i\\
\chi &=& \exp \left[ \sum_{\alpha} \frac{\lambda_{0}}{\omega_{\alpha}} \,
  (B_{\alpha}-B_{\alpha}^{\dagger}) \right]\, .
\ee
The new Hamiltonian  reads 
(with $\chi\chi^{\dagger} = \chi^{\dagger}\chi=1$)
\begin{align}
\bar{H} &= \sum_i (\epsilon_i-\Delta) a_i^{\dagger}a_i-\sum_{i,j; i\neq j} t_{ij} a_i^{\dagger} a_j\nonumber\\
& + \sum_{r,k,i} \left[ t_{ik}^{r}c_{kr}^{\dagger}a_i \chi +t_{ik}^{r*}a_i^{\dagger} \chi^{\dagger} c_{kr} \right]+H_{\rm L} +H_{\rm R} \nonumber\\
& + \sum_{\alpha} \omega_{\alpha} B_{\alpha}^{\dagger}B_{\alpha}+ \sum_{\alpha} \sum_{i,j; i \neq j} \lambda_{ij} \, a_i^{\dagger} a_j  (B_{\alpha}+B_{\alpha}^{\dagger})\\
\Delta &=\int d\omega D(\omega) \frac{\lambda_{0}^2}{\omega} \; .
\end{align}
Here we neglected terms with vibron-mediated electron-electron interaction
\cite{Boettger85}.
This is a reasonable approximation for the low hole density in DNA.
The purpose of the Lang-Firsov transformation is to remove the 
local electron-vibron
coupling term from the transformed Hamiltonian in exchange for the
transformed operators and the so-called polaron shift $\Delta$, 
describing the lower on-site energy of the polaron as compared to the
bare electron. However, the non-local coupling term
remains unchanged and has to be dealt with in a different way than the
local term (see below). There is an additional electron-vibron coupling 
due to the vibron shift generator $\chi$ in the transformed tunnel 
Hamiltonian from the leads. In this study we neglect effects arising
from this additional coupling. This is a valid approximation 
for $\Gamma^{\rm L,R} \gg \lambda_0 $ and the usual approximation 
taken in the literature \cite{Galperin05a,Gutierrez06}.

We introduce the retarded electron Green function
 \be
 G_{kl}^{\mathrm{ret}}(t)&=&-i\theta (t) \left\langle \left\lbrace  
 a_k(t)\chi(t),a_l^{\dagger} \chi^{\dagger} \right\rbrace  \right\rangle \, ,
\ee
where the thermal average is taken with respect to the 
transformed Hamiltonian, which does not explicitely include the 
local electron-vibron interaction. 
By applying the equation of motion (EOM) technique we can derive a 
self-consistent calculation scheme for  $G_{kl}^{\mathrm{ret}}(t)$ 
(see Appendix \ref{sec:appendix}).
From the Green function obtained by this scheme we extract the physical 
quantities of interest, like the density of states and the current.
The EOM technique for an interacting system generates correlation
functions of higher order than initially considered,  resulting in a 
hierarchy of equations that does not close in itself. Therefore, an
appropriate  truncation scheme needs to be applied.  In our case, we close 
the hierarchy on the first possible level neglecting all higher order
Green functions beyond the one defined above. In particular, our
approximations are perturbative to first order in $\lambda_1$ (for details
see Appendix \ref{sec:appendix}), restricting our study to relatively weak 
non-local electron-vibron coupling strengths.

For a DNA chain with $N$ bases the density of states is
\be
A(E)=-\frac{1}{\pi N}\, \sum_{i=1}^N {\rm Im}\left\{
G_{ii}^{\mathrm{ret}}(E)\right\} \, .
\ee
In the wide-band limit, the retarded electrode self-energies are constant 
and purely imaginary: 
$\Sigma_{ij}^{\rm L}=i\Gamma^{\rm L} \delta_{i1}
\delta_{j1}$ and 
$\Sigma_{ij}^{\rm R}=i\Gamma^{\rm R} \delta_{iN} \delta_{jN}$.

We evaluate the current using the relation\cite{Meir92}
\begin{align}
I=\frac{ie}{h} \int &d\epsilon \Big(\mathrm{tr}\left\lbrace \left[ 
f_{\rm L}(\epsilon)\Gamma^{\rm L}-f_{\rm R}(\epsilon)\Gamma^{\rm R}\right] 
\left(G^{\mathrm{ret}}(\epsilon)-G^{\mathrm{adv}}(\epsilon) \right) 
\right\rbrace \nonumber \\
&+\mathrm{tr}\left\lbrace \left[ \Gamma^{\rm L}-\Gamma^{\rm R}\right] 
G^{<}(\epsilon) \right\rbrace \Big) \; ,
\end{align}
where $f_{\rm L}(\epsilon)$ and $f_{\rm R}(\epsilon)$ are the Fermi 
distributions in the left and right lead, respectively.

To compute the `lesser' Green function $G^{<}(\epsilon)$, we use the
relation \cite{Mahan}
\be
G^{<}(\epsilon)=G^{\mathrm{ret}}(\epsilon)\left[\Sigma^{\mathrm{L}<}+
\Sigma^{\mathrm{R}<}+\Sigma_{\rm vib}^{<} (\epsilon)\right]
G^{\mathrm{adv}}(\epsilon) \, .
\ee
While the lesser electrode self-energies, such as $\Sigma^{\mathrm{L}<}$,
can be determined easily  within the above approximation for 
any applied bias, we have to
approximate the behavior of the lesser self-energy due to the vibrons 
$\Sigma_{vib}^{<} $. Extending the known relation for the equilibrium 
situation we write
\be
\Sigma_{\rm vib}^{<} (\epsilon)= -f_{\mathrm{eff}}(\epsilon)\left[ 
\Sigma_{\rm vib}^{\mathrm{ret}}(\epsilon)-
\Sigma_{\rm vib}^{\mathrm{adv}}(\epsilon) \right] \, ,
\ee
with an effective electron distribution $f_{\mathrm{eff}}= [f_{\rm
  L}(\epsilon)+f_{\rm R}(\epsilon)]/2$,
multiplying the equilibrium expressions for 
$\Sigma_{\rm vib}^{\mathrm{ret}}\, , \Sigma_{\rm vib}^{\mathrm{adv}}$.
Combining all terms we obtain a concise expression for the current, which can
be separated into `elastic' and `inelastic' parts as
\be
I=\frac{2e}{h} \int d\epsilon \,  
\left[T_{\rm el}(\epsilon)+ T_{\rm inel}(\epsilon) \right]
\left[ f_{\rm L}(\epsilon)-f_{\rm R}(\epsilon)\right] \, ,
\label{eq:curr}
\ee
where we identify the `elastic' and `inelastic' transmission functions\cite{Galperin04,Viljas05}
\begin{align}
T_{\rm el}(\epsilon) &= 2 \, \mathrm{tr} \left\lbrace \Gamma^{\rm R}
G^{\mathrm{ret}} (\epsilon)\Gamma^{\rm L}G^{\mathrm{adv}}(\epsilon)
\right\rbrace \\
T_{\rm inel}(\epsilon) &= \frac{i}{4} \, \mathrm{tr}  
\{ (\Gamma^{\rm R} +\Gamma^{\rm L})  G^{\mathrm{ret}}(\epsilon) \nonumber \\
&\times \left[ \Sigma_{\rm vib}^{\mathrm{ret}}(\epsilon)- 
\Sigma_{\rm vib}^{\mathrm{adv}}(\epsilon) \right] 
G^{\mathrm{adv}}(\epsilon)\} \, .
\end{align}
Note that also the `elastic' transmission depends on
the effects of vibrons, since the self-consistent evaluation of the 
Green function is performed in the presence of vibrons and environment.  
The inelastic contribution can also be termed `incoherent', as
typically the electrons will leave the DNA at a lower energy than they
enter it. 

\section{Results}
\label{sec:results}

In this section we analyze the effect of vibrations on the electronic 
properties of DNA, i.e., we determine the density of states, 
the transmission and the current. As explicit examples we consider 
homogeneous and inhomogeneous DNA sequences of 26 base pairs in the 
presence of a single vibrational mode as described in the previous 
section. For simplicity, we couple the left and right electrodes 
symmetrically to the DNA, so 
$\Gamma^{\rm L} = \Gamma^{\rm R} \equiv \Gamma$, and we
choose $\Gamma = 0.1\,e\mathrm{V}$. We further assume that the bias
voltage $V_b$ drops symmetrically across both electrode-DNA interfaces.

\subsection{Homogeneous Poly-(GC) DNA}
\label{subsec:homo}
\begin{figure}
	\centering
	\includegraphics[width=7cm]{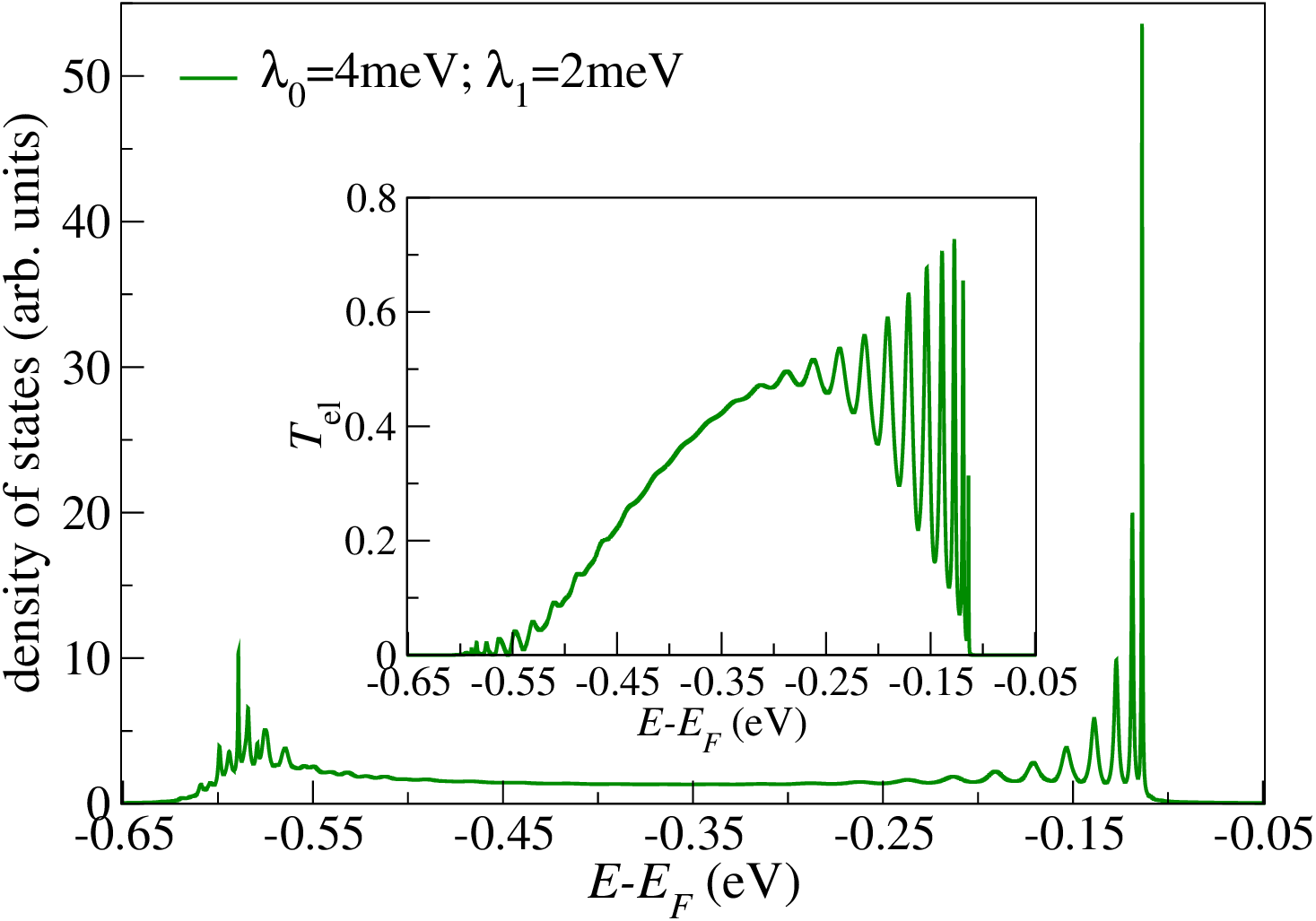}
	
\caption{(Color online) Density of states and transmission of Poly-(GC) with 26 base
  pairs and the following parameters: base pair on-site energy 
$\epsilon_G=-0.35\,e\rm V$, Fermi energy $E_F=0\,e\rm V$, vibrational energy $\hbar \omega_0=0.01\,e\rm V$,
cutoff $\hbar \omega_c=0.03\,e\rm V$, 
linewidth $\Gamma=0.1\,e\rm V$ and room temperature $k_B T=0.025\,e\rm V$.
The strong asymmetry of the curves with respect to the band center is
  a consequence of the non-local electron-vibron coupling $\lambda_1$.
\label{fig:G_trans2}}
\end{figure}

For a homogeneous DNA consisting of 26
Guanine-Cytosine base pairs we obtain a band-like density of states 
displayed in Fig.~\ref{fig:G_trans2}. With the
fairly small hopping element of $0.119\,e\mathrm{V}$ 
(see Tab.~\ref{tab:hopping}) for this finite system one can still
resolve the peaks due to single electronic resonances, especially near the
van-Hove-like pile up of states near the band edges. All states are
delocalized over the entire system. 
The inset displays the elastic  transmission,
showing that the states have a high  transmission of $T_{\rm el}\sim 0.5$,
with the states at the upper band edge showing the highest values. 
Both density of states and elastic transmission show a strong
asymmetry, which is a  direct consequence of the
non-local electron-vibron coupling in this model. 

\begin{figure}
	\centering
	\includegraphics[width=7cm]{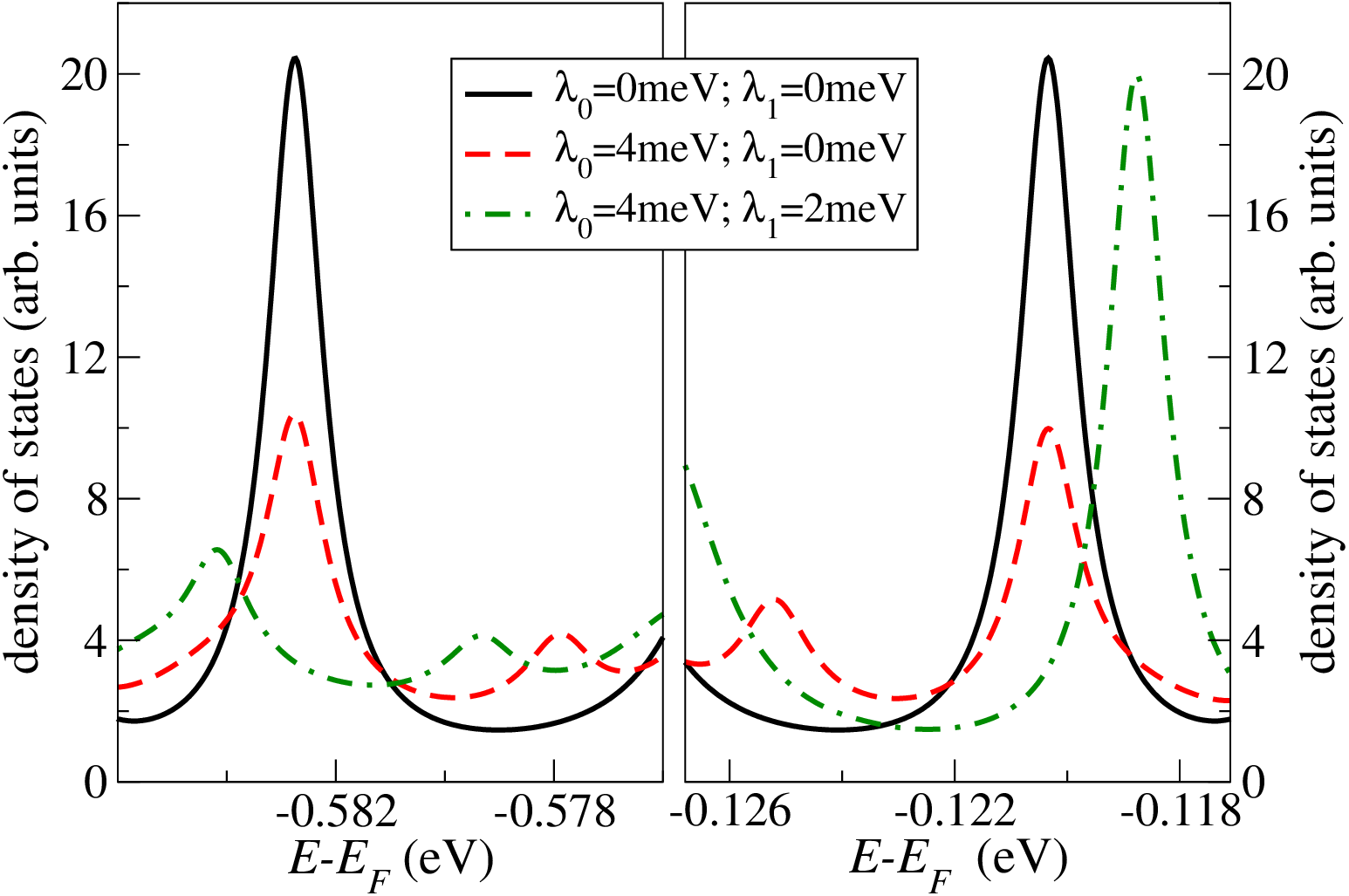}
\caption{(Color online) Density of states of Poly-(GC) with 26 base pairs and
  parameters as in Fig.~\ref{fig:G_trans2}.
The solid line shows the purely electronic resonances. Inclusion of
only a local electron-vibron coupling $\lambda_0$ reduces the weight at the
original electronic resonance in favor of `vibron satellites'
(dashed line). The addition of a  non-local electron-vibron coupling 
$\lambda_1$ (dash-dotted line) introduces shifts of the 
resonance peaks to the `outside' (changing the effective band width) 
as well as a strong asymmetry in the height of the resonances. 
}
\label{fig:G_M_Vergleich}
\end{figure}

To further elucidate this connection we take a closer look at the upper and
lower band edge of the density of states (see Fig.~\ref{fig:G_M_Vergleich}).
Without electron-vibron coupling (solid curve) we see the electronic 
resonances of equal height, positioned at the energies
corresponding to the `Bloch'-like states of this finite size 
tight-binding chain. If we include
only local electron-vibron coupling (dashed  line),  
vibron  satellite states appear, and the spectral weight of the original
electronic resonances decreases, consistent with the spectral sum rule.
Note that the displayed vibron satellites are not satellites of the 
displayed electronic states, but emerge from other states at higher
and lower energies. Indeed the difference in peak positions is
not equal to $\hbar \omega_0$.
Inclusion of the non-local coupling $\lambda_{1}$ shifts the original electronic
resonance positions (dashed-dotted line).  
In the present example, with positive sign of
$\lambda_{1}$,  the resonances are shifted to the `outside', corresponding to an 
effective increase in bandwidth; for the opposite sign of $\lambda_{1}$ the resonances
shift to the `inside'. Furthermore, a distinct asymmetry of the resonances
is observed, i.e. the upper band edge states have a larger peak height 
than the lower band edge states. This asymmetry in the density of
states comes with a corresponding asymmetry in the elastic transmission, see
Fig.~\ref{fig:G_trans2} for the overall view.

\begin{figure}
	\centering
	\includegraphics[width=7cm]{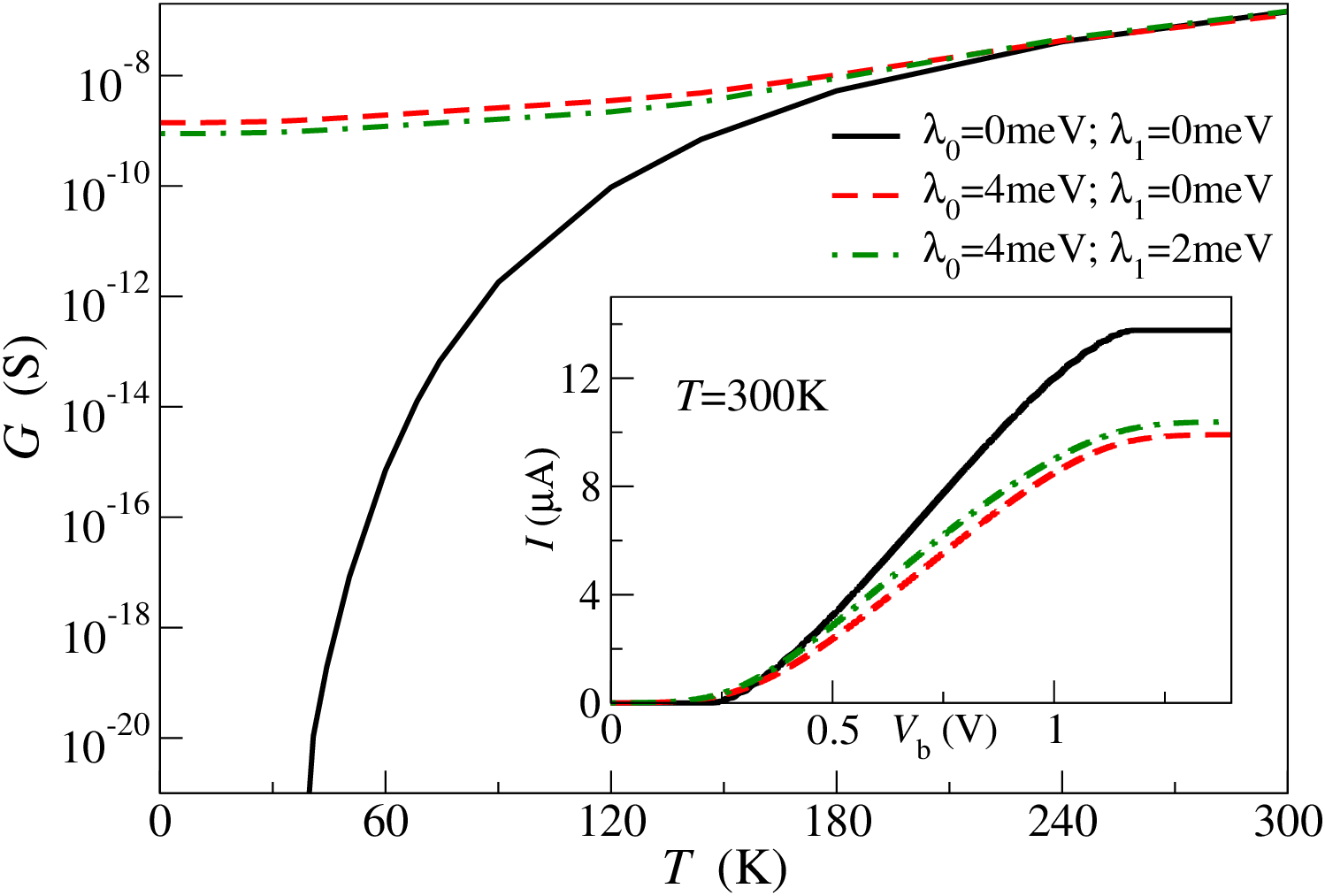}
	\caption{(Color online) Zero-bias conductance and {$I$-$V$}-characteristics for
	  Poly-(GC) with 26 base pairs and parameters as in 
	Fig.~\ref{fig:G_trans2}. The inclusion of vibrons increases
	  the zero-bias conductance at low temperatures ($k_{\rm B} T$ roughly 
	  below $\hbar \omega_0$) by several orders of magnitude. At room
	  temperature, however, the zero bias conductance is slightly reduced.
	  Inset: the {$I$-$V$}-characteristics shows a 
	  `semiconducting' behavior at 
	  room temperature. The non-local electron-vibron coupling 
	  $\lambda_{1}$ increases both the non-linear
	  conductance in the gap and around the threshold, leading to
	  a slightly enhanced current. 
	}
	\label{fig:G0cond_Temp}
\end{figure}
As shown in Fig.~\ref{fig:G0cond_Temp} the coupling to vibrons
strongly increases the zero-bias conductance at
low temperatures, whereas at high temperatures the conductance
slightly decreases (dashed and dash-dotted line). 
This effect has been observed before, e.g. in Ref.~\onlinecite{Gutierrez05b}.
At low temperatures, the conductance is increased since the density of
states at the Fermi energy is effectively enhanced due to (broadened) 
vibronic `satellite' resonances. The transport remains `elastic', 
i.e. electrons enter and leave
the DNA at the same energy (first contribution to the current 
Eq.~\ref{eq:curr}). At sufficiently high temperatures, however, the 
back scattering of electrons due to vibrons reduces the conductance in
comparison to situation without electron-vibron coupling (solid line).
       
The inset of Fig.~\ref{fig:G0cond_Temp} shows a 
typical {$I$-$V$}-characteristic for the system. 
A quasi-semiconducting behavior is observed, where the size of the
conductance gap is determined by the energetic distance of the 
Fermi energy to the (closest) band edge. 
After crossing this threshold, the current
increases roughly linear with the voltage until at larger bias it
saturates when the right chemical potential drops below the lower
transmission band edge. Small step-like wiggles due to 
the `discrete' electronic states are visible at low temperature (not
shown), but are smeared out at room temperature. 
The current is dominated by the elastic transmission, as expected 
for a homogeneous system. 

The non-local coupling has a quantitative 
effect on the nature of the {$I$-$V$}-curve. The zero bias conductance as
well as the non-linear conductance around the threshold are increased
by close to a factor  $1.2$. 
This increase is directly related to the 
enhancement of the density of states and elastic transmission 
around the upper band edge
(see Figs.~\ref{fig:G_trans2} and \ref{fig:G_M_Vergleich}). 

\subsection{Inhomogeneous DNA}
\label{subsec:inhomo}
Inhomogeneous DNA sequences show a transport behavior which differs
significantly from that of the homogeneous Poly-(GC) sequence.
As a specific example, we analyze the sequence 
5'-CAT TAA TGC TAT GCA GAA AAT CTT AG-3' (plus complementary strand), 
which has been investigated experimentally by Porath et al.\cite{Cohen05}. 
\begin{figure}
	\centering
	\includegraphics[width=7cm]{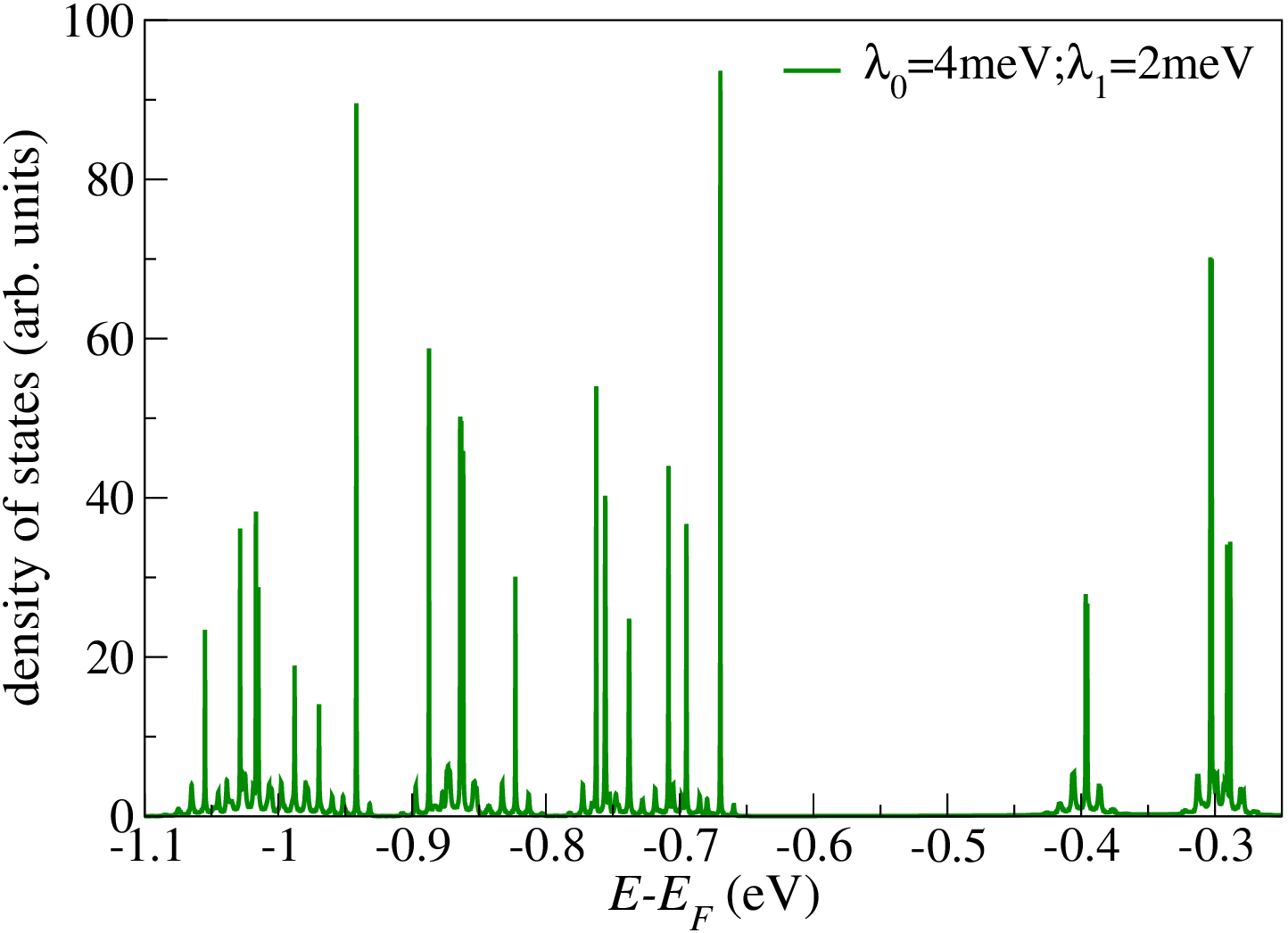}
	\caption{(Color online) Density of states of an inhomogeneous DNA with
	  sequence (5'-CAT TAA TGC TAT GCA GAA AAT CTT AG-3').
	 We chose the following parameters: 
	 GC on-site energy $\epsilon_G=-0.35\,e\rm
	  V$, AT  on-site energy $\epsilon_A=-0.86\,e\rm V$ , 
	  Fermi energy $E_F=0\,e\rm V$, vibron energy 
	  $\hbar \omega_0=0.01\,e\rm V$, cutoff $\hbar \omega_c=0.03\,e\rm V$,
	 linewidth $\Gamma=0.1\,e\rm V$ and room temperature $k_B
	  T=0.025\,e\rm V$.
	  The density of states is fragmented into `bunches' of
	  strongly localized states with very low elastic transmission. 
}  
	  \label{fig:G_Porath}
\end{figure}
The density of states is displayed 
in Fig.~\ref{fig:G_Porath}. Rather than traces of bands it 
now shows discrete `bunches' of states due to the disorder in the sequence.
All states are strongly localized, extending over at
most a few base pairs \cite{Klotsa05b}. 
The right-most (largest energy) bunch of states is due to the GC
base pairs. Two of these GC pairs are the only base pairs that are 
directly coupled to the metallic electrodes. 
Note that the equilibrium Fermi level is set at $E_F=0 \,e\rm V$, 
roughly $0.35\,e\rm V$ above these states. The first states with
mostly AT character are located around $ -0.7\,e\rm V$.

As to be expected the elastic transmission through these
localized states is extremely low. 
The largest contribution to the elastic transmission stems from 
the AT-like states around an energy $\epsilon_A=-0.86\,e\rm V$
(note that the considered sequence is AT rich).
But even these states have an elastic transmission of less than
$10^{-14}$ for the parameters we use.
Consequently, the `elastic' quasi-ballistic transmission of
electrons is completely negligible for the considered sequence.

\begin{figure}
	\centering
	\includegraphics[width=7cm]{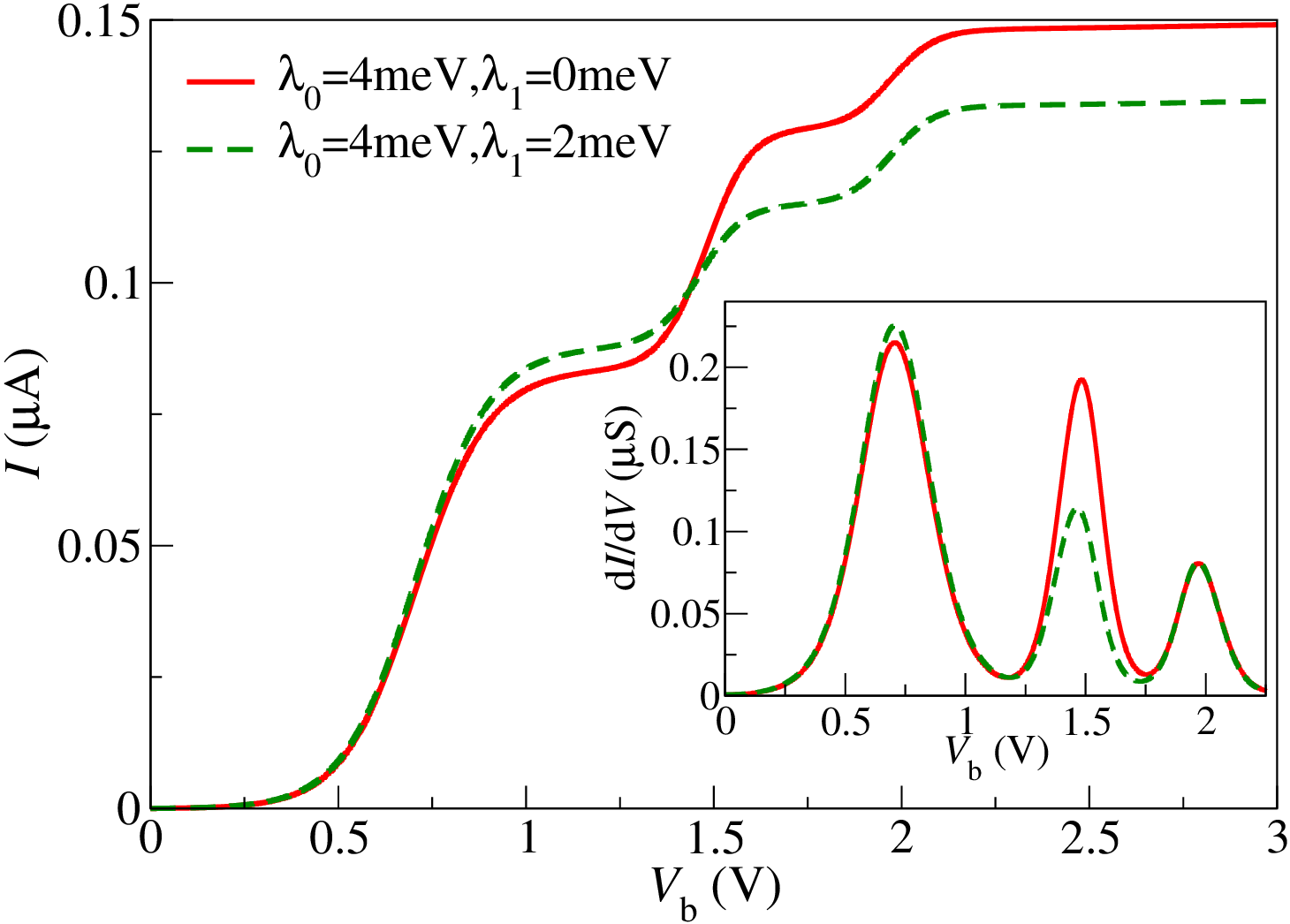}
	\caption{(Color online) {$I$-$V$}-characteristics and differential conductance 
	for an inhomogeneous DNA with sequence 
	(5'-CAT TAA TGC TAT GCA GAA AAT CTT AG-3'). Parameters are the
	same as in Fig.~\ref{fig:G_Porath}. The inclusion of a
	non-local electron-vibron coupling $\lambda_1$ leads to
	changes in the conductance, depending on the
	nature of the relevant state.
}
\label{fig:IV_M1vergleich} 
\end{figure}

In spite of the localization of the electron states, a rather significant
current can be transmitted, as displayed in
Fig.~\ref{fig:IV_M1vergleich}. It is due to the inelastic 
contributions to transport, where electrons dissipate (or absorb)
energy during their motion through the DNA. 
Roughly speaking,
the transported electrons excite the vibrons which in turn either 
dissipate their energy to the environment or `promote' 
other electrons, thus increasing their probability to hop 
to neighboring but energetically distant base pairs.
This inelastic transmission strongly depends 
on the specific states (in contrast to the band-like transmission 
for the homogeneous sequence). As a consequence, the inelastic transmission of
different states can differ by  several orders of magnitude.
Together with the bunched density of states this leads
to the step-like behavior for the current displayed in 
Fig.~\ref{fig:IV_M1vergleich}. The first step centered 
around $V_b \sim 0.7\, \rm V$ roughly corresponds to 
states with GC character, whereas the second step corresponds to
states with mixed AT-GC character at  $-0.7\,e\rm V$. Here, the
GC states display a larger inelastic transmission as can be
seen from the large non-linear conductance peak around 
$V_b \sim 0.6-0.7\, \rm V$ (see inset of Fig.~\ref{fig:IV_M1vergleich}). 

The non-local electron-vibron coupling $\lambda_1$
for this sequence leads  to qualitative change of
the {$I$-$V$}-characteristics, depending on the details of the 
nature of the states and therefore explicitly on the DNA sequence. 
The current on the lowest bias
plateau is increased relative to the case with only local
electron-vibron coupling, although the GC states do barely shift
towards the Fermi energy.
However, the inelastic transmission of 
the states is slightly increased (see inset), leading to an 
increased current on the first plateau (dashed line).

In contrast, the conductance due to states 
with mixed AT-GC nature is much reduced
(almost by a factor of two, see middle peak in the
inset of Fig.~\ref{fig:IV_M1vergleich}) which leads to a smaller
increase of the current for the middle step. 
Obviously, the transmission of these
mixed states is reduced by the `vibron assisted electron hopping'.
On the other hand, the last step at $\sim 2 V$ is almost unaffected.
   
While the changes of the {$I$-$V$}-characteristics due to non-local
electron-vibron coupling are relatively small for the
present sequence and model parameters, the observed sensitivity of 
the inelastic transmission suggests that other sequences could display 
much larger effects. Furthermore, quantum chemistry calculations 
\cite{Starikov05} suggest that the 
local and non-local electron-vibron couplings can be of the order of 
$\sim 10\, \mathrm{m}e \rm V$, i.e. larger than what we considered here. 
Inhomogeneities in the electron-vibron coupling, not covered in the 
present calculation, might have a further impact. 

The DNA sequence we considered was  investigated in transport
experiments, and we should compare the experimental and theoretical 
results. As some important factors are still not well determined, 
a quantitative comparison is not feasible. However, we observe both in 
 experiment and theory roughly a `semiconducting' 
{$I$-$V$}-characteristics with (sometimes) steplike features. The size of the 
currents is roughly comparable, of the order 
of $\sim 80\, \mathrm{nA}$ at a bias of  $V_b =1\, \mathrm{V}$. 
As the choice of the position of the Fermi energy defines the size of the
`semiconducting' gap, this gap could be adjusted to fit the experiment.  
On the other hand, the value of the current for this sequence (with parameters 
derived from quantum chemistry calculations)
can not be simply scaled by changing a single `free' parameter like the
electrode-DNA coupling $\Gamma$. 

\begin{figure}
	\centering
	\includegraphics[width=7cm]{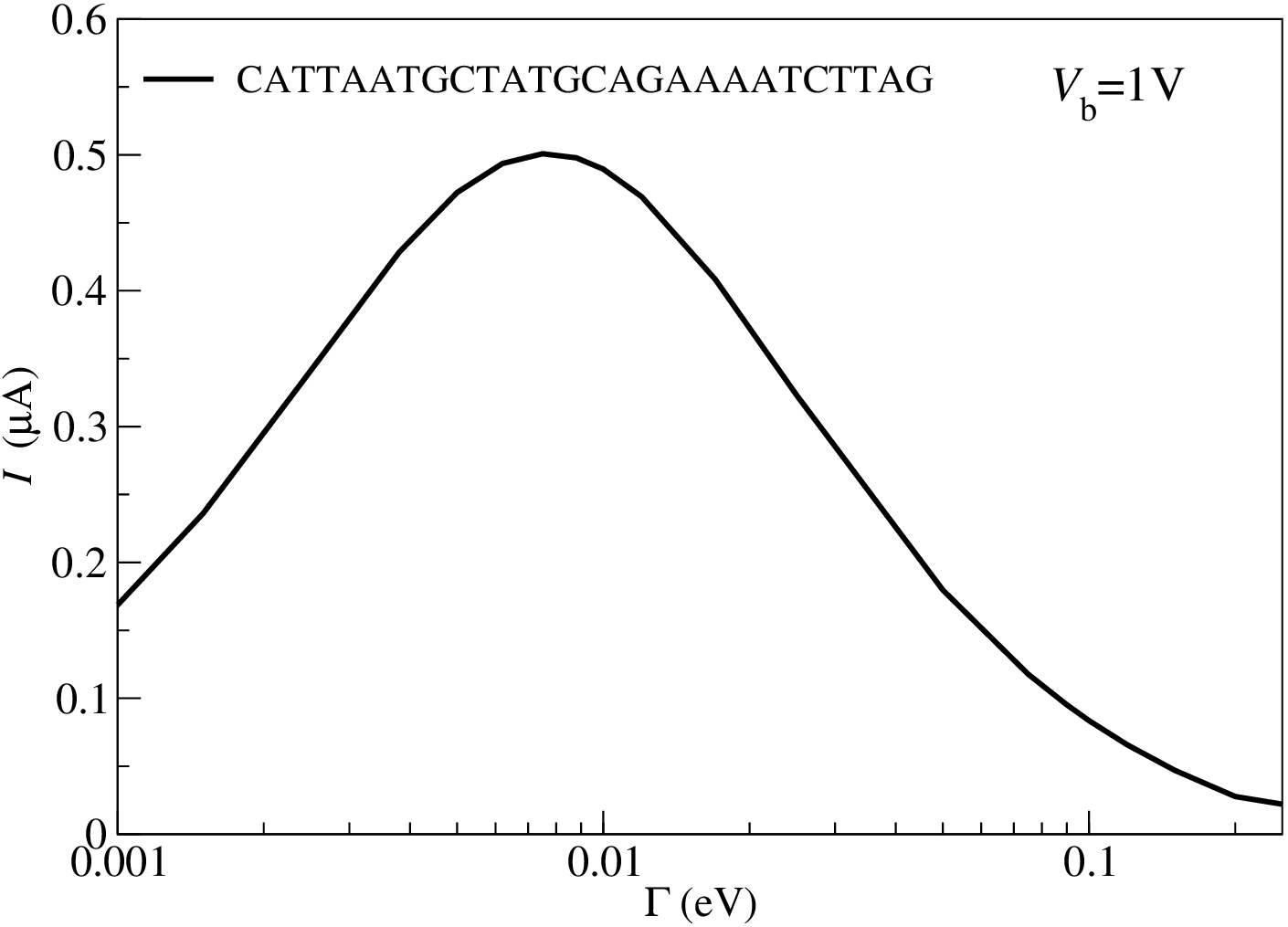}
	\caption{Current at a bias of $V_b= 1\, \mathrm{V}$ as a function
	of electrode-DNA coupling  $\Gamma$
	for the inhomogeneous DNA with sequence 
	(5'-CAT TAA TGC TAT GCA GAA AAT CTT AG-3'). Other parameters are the
	same as in Fig.~\ref{fig:G_Porath}. The current is a
	non-monotonous function of  $\Gamma$ and peaks around a value  
	$\Gamma_{max}$ where the imaginary part of the vibron self
	energy $\Sigma_{\rm vib}$ is of the same size as $\Gamma$.
	}
\label{fig:gamma_dep} 
\end{figure}

For the case of the
homogeneous sequence, the current at a given bias 
(say, at $V_b= 1\, \mathrm{V}$) grows
monotonically with increasing $\Gamma$ 
(as long $\Gamma$ is smaller than the hopping amplitude
$t_{ij}$), as is expected from quasi-ballistic Landauer-type transport.
In contrast, for the inhomogeneous sequence, 
the current is a {\it non-monotonic} function of $\Gamma$, 
see Fig.~\ref{fig:gamma_dep}. 
In particular, the current at the first plateau (at $V_b= 1\, \mathrm{V}$) 
initially grows as we decrease  $\Gamma$ from the value used 
in the above figures  ($\Gamma =0.1\,e\mathrm{V}$), 
up to a point at which the imaginary part of
the vibron self energy $\Sigma_{\rm vib}$ is of the same size as
$\Gamma$. This happens around $\Gamma_{max} \sim 0.01\,e\mathrm{V}$.
The current at $\Gamma_{max}$ is of order of $\sim 500\, \mathrm{nA}$. 
If $\Gamma$ is decreased further, the current drops rapidly from the
maximal value.\footnote{Note that our assumption $\Gamma >> \lambda_0$
breaks down at some point. Nevertheless, the decrease of the current
at very small $\Gamma$ makes  physical sense.}
On the other hand, if $\Gamma$ is increased above the value
$\Gamma =0.1\,e\mathrm{V}$, the current also drops initially, before at
very large $\Gamma$ quasi-ballistic transport becomes dominant and the
current increases again (not shown in the figure). 

Summarizing these results, 
we conclude that for the given model parameters, i.e. for values of 
$\Gamma$ in the large range $1- 200\, \mathrm{m} e \rm V$, likely to be 
realistic for present-days transport experiments in DNA,
the current at the first plateau lies in the range of $50- 500\, \mathrm{nA}$. 

\section{Summary}
\label{sec:summary}
To summarize, we have presented a technique that allows the computation
of electron transport through short sequences of DNA,
including local and non-local coupling to vibrations and a dissipative
environment. Using an equation-of-motion approach we identify 
elastic and inelastic contributions to the current. For homogeneous
DNA sequences, the transport is dominated by elastic quasi-ballistic 
contributions through a band-like density of states
(Fig.~\ref{fig:G_trans2},\ref{fig:G_M_Vergleich}),
which display an asymmetry due to the non-local electron-vibron
coupling. The coupling to vibrations strongly enhances  the 
zero-bias conductance at low temperatures. The current at finite bias
above the `semiconducting' gap, however, is only quantitatively
modified by  the non-local electron-vibron coupling 
(Fig.~\ref{fig:G0cond_Temp}). 
For inhomogeneous DNA sequences, the transport is almost entirely due
to inelastic processes, the effectiveness of which is 
strongly sequence dependent (Fig.~\ref{fig:G_Porath}). 
For the considered example sequence
the non-local electron-vibron coupling qualitatively modifies the 
{$I$-$V$}-characteristics (Fig.~\ref{fig:IV_M1vergleich}). We also point out
that the current through inhomogeneous DNA sequences depends
non-monotonically on the  electrode-DNA coupling $\Gamma$ 
(Fig.~\ref{fig:gamma_dep}).

{\em Acknowledgments.} We acknowledge stimulating discussions with 
J. Viljas,  J. Starikov, A.-P. Jauho, E. Scheer and
W. Wenzel. We also thank the Landesstiftung Baden-W\"urttemberg for
financial support via the Kompetenznetz ``Funktionelle
Nanostrukturen''.

%\bibliographystyle{apsrev}
%\bibliography{thesis}

\begin{appendix}
\section{Equation of motion}
\label{sec:appendix}
Before applying the equation of motion we separate the retarded 
electron Green function into two parts,
\be
G_{kl}^{\mathrm{ret}}(t)&=&-i\theta (t) \left\langle \left\lbrace  
a_k(t)\chi(t),a_l^{\dagger} \chi^{\dagger} \right\rbrace  \right\rangle \nonumber \\
&=&\underbrace{-i\theta (t) \left\langle a_k(t)\chi(t)
a_l^{\dagger}\chi^{\dagger} \right\rangle}_{G_{kl}^{(1)}(t)} \nonumber \\
& & \underbrace{-i\theta (t) \left\langle a_l^{\dagger}\chi^{\dagger} 
a_k(t)\chi(t) \right\rangle}_{G_{kl}^{(2)}(t)} \, .
\label{eq:G_def}
\ee
This is necessary, because for $G_{kl}^{(1)}(t)$ and
$G_{kl}^{(2)}(t)$ self-consistency equations can be 
derived via the equation-of-motion technique (EOM) 
(The equation of motion applied to the retarded Green function $G_{kl}^{\mathrm{ret}}(t)$ 
leads to an equation containing not only the retarded Green function). 
The EOM technique for an interacting system generates a hierarchy of correlation
functions that does not close in itself. Therefore, an
appropriate  truncation scheme needs to be applied.  Here we close 
the hierarchy at the first possible level, i.e. we neglect all higher order
Green functions beyond the one defined above. 

From the equation of motion we obtain the following 
expression for $G_{kl}^{(1)}(t)$ defined in
Eq.~(\ref{eq:G_def})
\begin{align}
\lefteqn{\sum_{j} \left[ (i \frac{\partial}{\partial t}-\epsilon_k)\delta_{jk}+t_{kj} \right]\, G_{jl}^{(1)}(t) }  \nonumber \\
&= \delta(t) \left\langle a_k a_l^{\dagger} \right\rangle +i\theta (t) 
\Delta \left\langle a_k(t)\chi(t)a_l^{\dagger}\chi^{\dagger} \right\rangle  \nonumber \\
&-i\theta (t) \left\{ \sum_{j \neq k, \alpha} \lambda_{kj} \, \left\langle a_j(t)   \left[ B_{\alpha}(t)+B_{\alpha}^{\dagger}(t)\right]  \chi(t) a_l^{\dagger}\chi^{\dagger} \right\rangle \right . \nonumber\\
&+ \sum_{\alpha} \lambda_{0} \, \left\langle a_k(t) B_{\alpha}(t) \chi(t) a_l^{\dagger}\chi^{\dagger} \right\rangle \nonumber\\
&+ \sum_{\alpha} \sum_{i,j;j \neq i}  \frac{ 2\, \lambda_{ij} \, \lambda_{0}}{\omega_{\alpha}} \left\langle a_k(t) a_i^{\dagger}(t)   a_j(t) \chi(t) a_l^{\dagger}\chi^{\dagger} \right\rangle \nonumber\\
&+ \sum_{\alpha} \lambda_{0} \, \left\langle a_k(t)  \chi(t) B_{\alpha}^{\dagger}(t) a_l^{\dagger}\chi^{\dagger} \right\rangle + \left . \sum_n V_{nk}^{*} \left\langle c_n(t) a_l^{\dagger}\chi_l^{\dagger} \right\rangle \right \} 
\end{align}
and a similar relation for  $G_{kl}^{(2)}(t)$.

The expressions $\left\langle a_j(t)  B_{\alpha}(t) \chi(t) a_l^{\dagger} \chi^{\dagger} \right\rangle$ 
and similar higher order correlation function are approximated by assuming
\be
\left\langle a_j(t)  B_{\alpha}(t) \chi(t) a_l^{\dagger}
\chi^{\dagger} 
\right\rangle_{\bar H} \approx F_{\alpha}(t) \left\langle a_j(t) \chi(t) a_l^{\dagger} 
\chi^{\dagger} \right\rangle_{\bar H} .
\ee
The function $F_{\alpha}(t)$ is obtained by considering a Hamiltonian $H_0$ 
without electron-vibron coupling
and calculating the same higher order correlation function $\left\langle a_j(t)  B_{\alpha}(t) \chi(t) a_l^{\dagger} 
\chi^{\dagger} \right\rangle_{H_0}$, 
where now the average is taken with respect to $H_0$.
Then the electronic and vibronic correlators factorize,
\begin{align}
\left\langle a_j(t)  B_{\alpha}(t) \chi(t) a_l^{\dagger} \chi^{\dagger} \right\rangle_{H_0}
=\left\langle a_j(t) a_l^{\dagger} \right\rangle_{H_0^{\rm el}} \left\langle B_{\alpha}(t) \chi(t)  
\chi^{\dagger} \right\rangle_{H_0^{\rm vib}} \, ,
\end{align}
where $H_0^{\rm el}$ and $H_0^{\rm vib}$ are the electronic and 
vibronic parts of $H_0$.

After some straight-forward algebra (cf Ref.~\onlinecite{Boettger85}) we obtain
\be
\left\langle B_{\alpha}(t) \chi(t) \chi^{\dagger} \right\rangle_{H_0^{\rm vib}}=F_{\alpha}(t) \left\langle 
\chi(t) \chi^{\dagger} \right\rangle_{H_0^{\rm vib}}
\ee
 and consequently 
\be
\left\langle a_j(t)  B_{\alpha}(t) \chi(t) a_l^{\dagger}
\chi^{\dagger} 
\right\rangle_{H_0}=F_{\alpha}(t) \left\langle a_j(t) \chi(t) a_l^{\dagger} 
\chi^{\dagger} \right\rangle_{H_0} \, .
\ee
Because the strength of the electron-vibron coupling in $\bar H$ is 
proportional to $\lambda_1$, 
this approximation is valid for not too large values of $\lambda_1$.

Expressions like $ \left\langle a_l^{\dagger}\chi^{\dagger} a_k(t) a_i^{\dagger}(t)   a_j(t) \chi(t)  \right\rangle$ 
are treated in a mean-field like manner:
\be
\lefteqn{\left\langle a_k(t) a_i^{\dagger}(t)  a_j(t) \chi(t) a_l^{\dagger}\chi^{\dagger} \right\rangle} \nonumber\\
& \approx & \left\langle a_k(t) a_i^{\dagger}(t) \right\rangle \left\langle a_j(t) \chi(t) a_l^{\dagger}\chi^{\dagger} \right\rangle \nonumber \\
&-& \left\langle a_j(t) a_i^{\dagger}(t) \right\rangle \left\langle a_k(t) \chi(t) a_l^{\dagger}\chi^{\dagger} \right\rangle \, .
\ee

Using the above approximations we obtain after Fourier 
transformation and crossover to the continuous sprectrum
\begin{eqnarray}
\lefteqn{\sum_{j} \left[ (E-\epsilon_k)\delta_{jk}+t_{kj} \right]\, G_{jl}^{(1)}(E)  }  \nonumber \\
&&= \left\langle a_k a_l^{\dagger} \right\rangle -\Delta G_{kl}^{(1)}(E) \nonumber \\
&&+\int d\omega D(\omega) \Bigg\lbrace -\sum_i \sum_{j \neq i}  \left\langle a_j a_i^{\dagger} \right\rangle \frac{2\lambda_{ij} \lambda_{0}}{\omega}  G_{kl}^{(1)}(E) \nonumber\\
&&+\sum_i \sum_{j \neq i}  \left\langle a_k a_i^{\dagger} \right\rangle \frac{2\lambda_{ij} \lambda_{0}}{\omega}  G_{jl}^{(1)}(E) \nonumber\\
&&+\sum_{j \neq k} \frac{\lambda_{kj} \lambda_{0}}{\omega}
\left[ \int dt e^{iEt}\left[ F_{1}(t,\omega) - 1 \right]\, G_{jl}^{(1)}(t) \right]  \nonumber\\
&&+ \frac{\lambda_{0}^2}{\omega} \left[ \int dt e^{iEt}F_{1}(t,\omega) \, G_{kl}^{(1)}(t) \right] \Bigg\rbrace \nonumber\\
&&+\sum_j \Sigma_{kj}^{\rm L} G_{jl}^{(1)}(E)+\sum_j \Sigma_{kj}^{\rm R} G_{jl}^{(1)}(E) 
\label{eqn:G1}
\end{eqnarray}
with
\be
F_1(t,\omega) &=& \left(N(\omega)+1 \right) e^{-i\omega t}- N(\omega) e^{i\omega t} \, ,
\ee
and $\Sigma_{\rm R/L}$ are the right and left electrode self-energies.
A similar relation holds for  $G_{kl}^{(2)}(E)$.

We can now identify 
\be
(E-\epsilon_k)\delta_{jk}+t_{jk}+i0^+=\left[ G_0^{\mathrm{ret}}(E) \right]_{jk}^{-1} \, ,
\ee
where $G_0^{\rm ret}(E)$ is the retarded Green 
function for the isolated DNA without electron-vibron interaction. 
The validity of this equation can easily be seen by computing the 
equation of motion for $G^{\mathrm{ret}}(t)$ for the isolated DNA 
without electron-vibron coupling.

\end{appendix}

\end{document}